# Uncovering the (un-)occupied electronic structure of a buried hybrid interface

S. Vempati[1,*], J.-C. Deinert[1,†], L. Gierster[1‡], L. Bogner[1], C. Richter[1,§], N. Mutz[2], S. Blumstengel[2], A. Zykov[2], S. Kowarik[2], Y. Garmshausen[3], J. Hildebrandt[3], S. Hecht[3], J. Stähler[1]

[1] Fritz-Haber-Institut der Max-Planck-Gesellschaft, Abt. Physikalische Chemie, Faradayweg 4-6, 14195 Berlin, Germany

[2] Humboldt Universität zu Berlin, Institut für Physik & IRIS Adlershof, Newtonstr. 15, 12489 Berlin, Germany

[3] Humboldt Universität zu Berlin, Institut für Chemie & IRIS Adlershof, Brook-Taylor-Str. 2, 12489 Berlin, Germany

**Abstract**

**The energy level alignment at organic/inorganic (o/i) semiconductor interfaces is crucial for any light-emitting or -harvesting functionality. Essential is the access to both occupied and unoccupied electronic states directly at the interface, which is often deeply buried underneath thick organic films and challenging to characterize. We use several complementary experimental techniques to determine the electronic structure of p-quinquephenyl pyridine (5P-Py) adsorbed on ZnO(10-10). The parent anchoring group, pyridine, significantly lowers the work function by up to 2.9 eV and causes an occupied in-gap state (IGS) directly below the Fermi level $E_F$. Adsorption of upright-standing 5P-Py also leads to a strong work function reduction of up to 2.1 eV and to a similar IGS. The latter is then used as an initial state for the transient population of three normally unoccupied molecular levels through optical excitation and, due to its localization right at the o/i interface, provides interfacial sensitivity, even for thick 5P-Py films. We observe two final states above the vacuum level and one bound state at around 2 eV above $E_F$, which we attribute to the 5P-Py LUMO. By the separate study of anchoring group and organic dye combined with the exploitation of the occupied IGS for selective interfacial photoexcitation this work provides a new pathway for characterizing the electronic structure at buried o/i interfaces.**

[*] Current affiliation: Department of Physics, Indian Institute of Technology Bhilai, Raipur-492015, India
[†] Current affiliation: Helmholtz-Zentrum Dresden-Rossendorf, Bautzner Landstr. 400, 01328 Dresden, Germany
[‡] Corresponding author: gierster@fhi-berlin.mpg.de
[§] Current affiliation: Leibniz Institute of Surface Engineering (IOM), Permoserstr. 15, 04318 Leipzig, Germany

**Introduction**

Hybrid systems combining organic and inorganic semiconductors are promising candidates for optoelectronic and light-harvesting applications[1,2]. Optimization of the synergy of both materials in devices crucially depends on the electronic structure at the interface of the organic and inorganic counterparts. For example, efficient energy transfer and light emission for LED applications is observed upon aligning the band edges of the transparent and conductive metal oxide ZnO and a π-conjugated organic molecule[3]. In contrast, in hybrid solar cells, a type-II heterojunction is required to enable the dissociation of excitons generated by sunlight in the organic material, and the offset of the band edges is known to determine the charge injection barrier[4]. Yet, the energy level alignment (ELA) is not the only parameter influencing the charge separation efficiency across hybrid interfaces. Intense fundamental research over the last decades emphasized the importance of the wavefunction overlap between the frontier molecular orbitals (LUMO and HOMO) of the organic and the conduction or valence band of the inorganic material[5,6], the presence of interfacial states[7,8,9] and the role of the electron-hole interaction between the ionized organic molecule and the injected carriers[10,11,12,13]. Detailed understanding of the role of all these different contributions is crucial in order to proceed in device optimization, which calls for suitable model systems and adequate methods to analyze them.

The main difficulty in the determination of the ELA at o/i interfaces is the access to the electronic structure at the interface, which is buried underneath the organic material. In linear optical spectroscopies, usually, the bulk signals dominate over surface and interface contributions, and also non-linear optical techniques, which can be interface-sensitive[14,15,16], only provide information on optical resonances in the system. Another common approach is the use of photoelectron spectroscopy (PES) based on ultraviolet (UV) light or X-rays to determine the occupied electronic structure at interfaces of films with precisely controlled thickness[3,17,18]. However, these techniques do not provide information on the unoccupied electronic structure, and due to the limited mean free path of photoelectrons with high kinetic energies, they cannot access deeply buried interfaces.

One approach to circumvent these issues is two-photon photoelectron (2PPE) spectroscopy using ultrashort laser pulses. This technique allows the determination of absolute binding energies of occupied and unoccupied states of the system[19]. For instance, 2PPE spectroscopy has been successfully used to explore the electronic structure of metal- and semiconductor-organic interfaces[20,21,22,23,24,25]. Nevertheless, due to the longer mean free path of the low-kinetic energy photoelectrons in 2PPE spectroscopy, careful interpretation is needed, as both electronic structure and excited state lifetimes may vary as a function of film thickness[26].

Zinc oxide (ZnO) is a semiconductor with a wide band gap of 3.4 eV, which has a great potential as transparent conductive electrode in light-harvesting and opto-electronic applications[27,28]. Intrinsically n-doped, its bulk valence and conduction band extrema lie at 3.2 and 0.2 eV below and above the Fermi level $E_F$, respectively (cf. Fig. 1a)[29]. While the adsorption of atoms and molecules is known to induce interfacial in-gap states (IGS) at ZnO surfaces[30,31,32], different donor and acceptor adsorbates can vary the ZnO work function between 1.6 eV and 6.5 eV30[30,33,34]. IGS formation and work function reduction can occur even under UHV conditions by the adsorption of Hydrogen (H) from the residual gas[30]. It should be noted that also chromophores without particular donor or acceptor character can modify the work function of the substrate. For instance, p-sexiphenyl (6P) reduces ZnO work functions by 0.2 eV or even 0.85 eV when grown in an upright-standing or flat-lying fashion, respectively[17]. In a previous study[34], we combined the non-polar, mixed-terminated ZnO(10-10) surface with pyridine molecules, which, in the gas phase, exhibit a negative electron affinity of -0.62(5) eV[35], an intrinsic permanent dipole moment $\mu_{mol}$ of 2.2 D[36] and a vertical ionization potential of 9.6 eV[37]. As reproduced in the inset of Fig. 1(b), we observed a substantial work function reduction ΔΦ by up to 2.9 eV at a coverage of one monolayer (ML), down to Φ = 1.6 eV, which results from a combination of the pyridine dipole moment and the charge transfer induced by bond formation between the pyridine nitrogen atom and Zn atoms at the semiconductor surface[34]. Very recent theoretical work[38], furthermore, has shown that the optical spectrum of the pyridine/ZnO(10-10) system exposes several excitonic signatures that display hybrid character. Remarkably, the lowest-energy exciton exhibits a binding energy of 0.4 eV, almost seven times larger than the free exciton binding energy of bulk ZnO.

In this article, we present a novel, 2PPE-based approach to address the interfacial ELA at a model o/i interface. We characterize the occupied and unoccupied electronic structure of p-quinquephenyl pyridine (5P-Py) molecules adsorbed on ZnO(10-10) in three steps: (i) pure interfacial properties of pyridine/ZnO(10-10), (ii) upright-standing 5P-Py/ZnO(10-10) in the first monolayer, and (iii) flat-lying 5P-Py for multilayer coverages. Our experiments are complemented by photoluminescence (PL) and PL excitation (PLE) measurements of 5P-Py in solution and adsorbed on ZnO(10-10) as well as X-ray diffraction (XRD) experiments of multilayer samples. We find occupied, in-gap electronic states below $E_F$ for both, pyridine and 5P-Py, molecules adsorbed on ZnO(10-10) and a significant work function reduction that can be used to determine the coverage of the organic molecules. This occupied, interfacial IGS can be used to selectively populate unoccupied electronic levels in proximity to the interface. Consequently, interface sensitivity is reached even for 5P-Py multilayer coverages. At the interface, we identify three unoccupied states: One bound state, 2.0 eV above $E_F$, which we assign to the 5P-Py LUMO, and two states above the vacuum level $E_{vac}$ at $E - E_F$ = 3.4 eV and 4.1 eV, respectively. We show that the bound state at 2 eV is also present in the multilayer regime, however only weakly coupled to the ZnO CB. Beyond the comprehensive characterization of the electronic structure of a model o/i interface, this work demonstrates a proof-of-principle for a new approach to investigate the ELA at buried interfaces based on selective interfacial photoexcitation.

## 2. Experimental section

The molecule 5P-Py is a derivative of the technologically relevant chromophore 6P. Details on the synthesis can be found in the supporting information. In 5P-Py, the terminal carbon atom of 6P is replaced by a nitrogen atom so that one end group resembles the pyridine molecule, also with regard to the static dipole moment $\mu_{mol}$ along the molecular axis (see the inset in Figure 2(b)).

The samples were prepared and analyzed in a pair of coupled ultrahigh vacuum (UHV) chambers operating at base pressures on the order of $10^{-10}$ mbar and $10^{-11}$ mbar, respectively. The ZnO(10-10) surface (MaTecK GmbH) was cleaned by repeated cycles of Ar$^+$ sputtering (10 min, $p_{Ar}$ = 2.0 × 10$^{-6}$ mbar, 750 eV at 300 K) succeeded by 30 min annealing at 750-850 K (initial heating rate of 30 K/min) following established procedures[39].

Pyridine adlayers were prepared by exposing the freshly cleaned ZnO(10-10) substrate at 100 K to pyridine (Sigma Aldrich, 99.8%) vapor using a pinhole doser. The layer thickness was determined using thermal desorption spectroscopy (TDS) (cf. Ref. [34]) and given in monolayers (ML), which are used as a mass equivalent. The 5P-Py molecules were evaporated from a Knudsen cell (0.3-0.7 nm/min) onto the freshly cleaned ZnO(10-10) surface after careful degassing of the molecules at 10 K below their evaporation temperature of 540-550 K. Adsorption at a substrate temperature of 300 K led to the same spectral signatures as adsorption at 100 K and subsequent annealing to 300 K. A quartz crystal microbalance (QCM) was calibrated using atomic force microscopy (AFM) images of 5P-Py on sapphire and is used to determine the coverage. Here, nanometers (nm) refer to the thickness of a hypothetical smooth 5P-Py film and can be viewed as a mass equivalent. The adlayer preparation takes place before significant H adsorption from the residual gas in the UHV chamber can occur. Yet, H adsorption after preparation of the adlayer cannot be excluded. All the samples were transferred *in situ* into the analysis chamber, and photoemission measurements were carried out at a sample temperature of 100 K.

One-photon photoelectron (1PPE) and 2PPE spectroscopy experiments were performed using a regeneratively amplified femtosecond (~50 fs pulses) laser system (Ti:Sa, Coherent RegA, 200 kHz) which provides a fundamental photon energy of $h\nu_{IR}$ = 1.55 eV. A fraction of the total power was directed to an optical parametric amplifier (OPA) to generate photons of energies between 1.85 and 2.55 eV. Photons with an energy of 3.9 eV are generated by frequency doubling of the output of the OPA and 6.2 eV photons are obtained by frequency quadrupling the 1.55 eV energy photons. Spot sizes on the sample were on the order of 80x80 µm$^2$. The kinetic energy ($E_{kin}$) and intensity of the photoelectrons were detected using a hemispherical analyzer (PHOIBOS 100, SPECS GmbH) and the energy resolution of the experiment is smaller than 50 meV. The spectra were referenced to the Fermi level $E_F$ of the gold sample holder, which was in electrical contact with the sample. A bias voltage of – 2.5 to -5 V was applied to the sample with respect to the analyzer, enabling the detection of electrons of zero $E_{kin}$. The low-energy secondary electron cut-off ($E_S$) in the photoelectron spectra provides the work function ($\Phi$) of the surface by $\Phi = h\nu - (E_F - E_S)$.

In addition to one-color 2PPE, we performed two-color 2PPE. Here, two laser pulses (pump and probe) with different photon energies were spatially and temporally overlapped with the help of a delay stage to adjust the optical path length of the probe laser beam. The probe laser pulse was used to monitor the excited state population induced by the pump pulse at variable time delays. To show the transient excited state population, the delay-independent and uncorrelated background consisting of the 1PPE and 2PPE signals of both laser beams alone was subtracted. This background was determined from the signal at negative delays, i.e. for a pump-probe timing in which case the probe pulse arrives before the pump pulse.

PL, PLE, and XRD were measured *ex situ*. PL and PLE spectra of 5P-Py/ZnO(10-10) were obtained using a scanning PL spectrometer (Edinburgh Instruments, FLS 980) with a Xe-lamp as excitation source. The excitation as well as the emitted beams were each dispersed in a double monochromator and the emission was detected by a photomultiplier tube. PL was excited a 3.88 eV and PLE was detected at 2.85 eV. PL and PLE of 5P-Py in solution were performed in chloroform at 298 K on a Cary Eclipse Fluorescence Spectrophotometer. An excitation energy of 3.9 eV and an emission energy of 3.1 eV was used for PL and PLE, respectively.

XRD experiments were performed on a lab-based X-ray diffractometer at a Cu-*Kα* wavelength of $\lambda$ = 1.5406 Å from a rotating anode source. The measurements were carried out at room temperature in a high vacuum chamber. In our XRD experiment the intensity of the specularly reflected X-ray beam was monitored as a function of the angle of incidence $\Theta$ and the detector angle $2\Theta$. The corresponding scattering vector $q_z$, which is the momentum transfer along the surface normal, was calculated using $q_z = 4\pi/\lambda \cdot \sin \Theta$, where $\lambda$ is the wavelength of the X-rays.

**3. Results**

**3.1 Occupied in-gap, hybrid state induced by the pyridine anchoring group**

Before discussing the ELA at 5P-Py/ZnO(10-10) interfaces, we focus on the electronic structure of the pure anchoring group, pyridine, adsorbed on the same surface. As discussed in the introduction and shown in the inset of Fig. 1(b), pyridine adsorption significantly reduces the sample work function[34]. We use this work function shift to extrapolate the expected ELA for pyridine/ZnO(10-10) in Fig. 1(a). Clearly, based on this simplistic approach, no electronic levels are expected in the energetic region of the ZnO band gap.

Fig. 1(b) shows an exemplary 1PPE spectrum of a 0.65 ML pyridine adlayer on the ZnO(10-10) surface measured with a photon energy of 3.1 eV, exceeding the sample work function at this coverage. Subtraction of the secondary electron background reveals, in contrast to the above-described expectations, a signature of an occupied electronic state (IGS) 0.28(3) eV below $E_F$ with a full width at half maximum of 400 meV. This energy corresponds to a binding energy of 0.48 eV with respect to the bulk conduction band minimum of ZnO. The IGS state is absent for pristine ZnO and cannot be a purely molecular level, as pyridine frontier molecular orbitals are several eV away. We conclude that it is a *hybrid* state arising from the interaction of pyridine with the ZnO(10-10) surface and, thus, is located at the interface.[40]

Recent theoretical work on the pyridine/ZnO(10-10) interface offers a likely explanation for the origin of the IGS based on detailed calculations of the electronic structure in the $G_0W_0$ approximation[38]. Remarkably, the resulting single-particle band structure does not expose any molecular states below the ZnO-dominated conduction band minimum that could explain our observation of an occupied hybrid state directly below $E_F$, i.e. several 100 meV below the conduction band minimum of ZnO. However, by solving the Bethe-Salpeter equation for this system, Turkina et al. provide the optical absorption spectrum of pyridine/ZnO(10-10), which exposes a low-energy hybrid exciton resonance with an extraordinarily large binding energy of 400 meV[38]. This binding energy is in good agreement with the binding energy of the IGS found in our experiments and, moreover, sufficient to stabilize the electron of the two-particle excitation *below* the Fermi level of the system. It seems likely that the experimentally observed IGS originates from electrons bound to photoholes at the o/i interface that are, due to their binding energy below $E_F$, sufficiently long-lived to form a quasistationary state that appears as an occupied electronic state in the photoemission spectra, which were acquired at a high repetition rate (200 kHz).

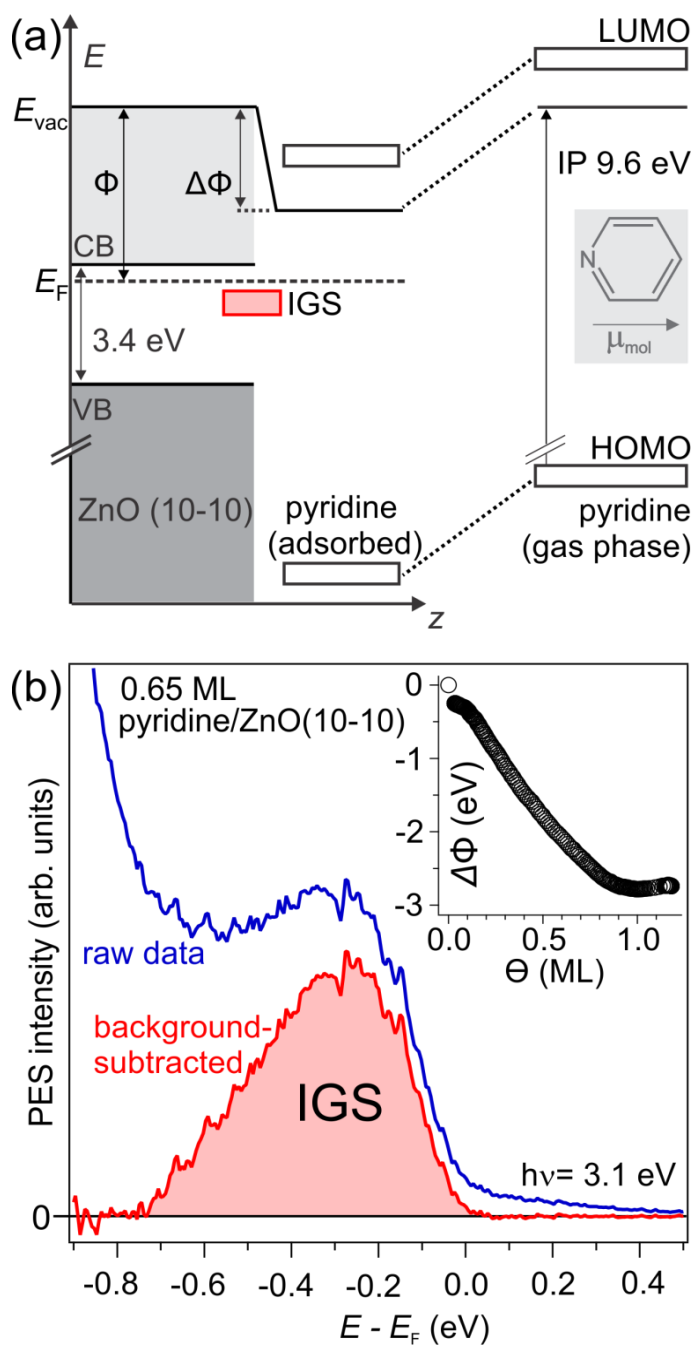

*Figure 1 (a) Schematic of the energy levels of pyridine in the gas phase (right) and the naively expected alignment at the pyridine/ZnO(10-10) interface (left) where the molecular levels shift rigidly due to vacuum level alignment and the build-up of an interfacial dipole. Inset: The pyridine molecule. (b) Photoelectron spectrum of the region around $E_F$ (blue). The combined system exhibits an IGS, which is highlighted after subtracting the secondary electron background of the 1PPE spectrum (red). Inset: Work function reduction $\Delta\Phi$ of up to -2.9 eV from the bare ZnO(10-10) to full monolayer (ML) coverage, reproduced from J. Chem. Phys. **139** 174701 (2013), with the permission of AIP Publishing.*

### 3.2 Characterization of the 5P-Py/ZnO(10-10) interface

The extension of pyridine to 5P-Py is expected to decrease the HOMO-LUMO gap (compare Fig. 1(a) and Fig. 2(a)) by increasing the delocalization within the π-conjugated system[41]. Fig. 2(b) presents PL and PLE measurements of 5P-Py solved in chloroform ($CHCl_3$) as well as in a film of 11 nm nominal thickness on ZnO(10-10) recorded at room temperature. Compared to the solution, the PLE is broadened but maximized at almost the same photon energy of 3.9 eV. The corresponding PL spectra display a Stokes shift and the 5P-Py film also exhibits a vibronic fine structure with a spacing of ≈ 0.15 eV, similar to 6P[42] (clearly discernable at low temperature PL).

We conclude that, as expected, the optical gap of 5P-Py is decreased with respect to the parent pyridine and amounts to 3.9 eV. Comparable PL and optical absorption properties can be found in the literature for 6P in solution[43] and in the solid state[42], also showing an enhanced Stokes shift in the condensed phase. The strong Stokes shift in phenylenes in solution has been attributed to different relative orientations (twisting) of the phenyl rings within one molecule in the ground and excited state[42,43], leading to a large molecular reorganization energy, while aggregation leads to the Stokes shift in the condensed phase. The asymmetry of absorption and emission properties of π-conjugated organic molecules with *para*-linked phenyl rings has been attributed to deviations from the mirror image symmetry of these compounds.[44]

Figure 2(c) presents a 1PPE spectrum of a 3 nm thick 5P-Py layer, deposited onto the cold ($T$= 100 K) ZnO(10-10) substrate (blue curve). Deposition leads to a work function reduction from 4.5 eV (bare ZnO(10-10)) to $\Phi_{5P-Py,cold}$ = 3.4(1) eV. While, for the deposition at 100 K substrate temperature, no sign of an in-gap state as for pyridine/ZnO(10-10) is observed, mild annealing of the film to 300 K for 10 minutes, leads to the build-up of an occupied state −0.7 eV relative to $E_F$, comparable to the IGS of the pure anchoring group adsorbed on ZnO(10-10). The spectral change close to $E_F$ is accompanied by a further work function reduction from $\Phi_{5P-Py,cold}$ = 3.4(1) eV to $\Phi_{5P-Py,annealed}$ = 2.8(1) eV. This significant work function reduction compared to the pristine ZnO surface of -1.7 eV is an order of magnitude larger than the $\Delta\Phi_{6P,upright}$ = -0.2 eV previously observed for upright-standing 6P/ZnO(000-1) and twice as large as $\Delta\Phi_{6P,flat}$ = -0.85 eV for flat-lying 6P/ZnO(10-10)[17], highlighting the important role of the pyridine anchoring group in binding 5P-Py to ZnO. As sketched in the inset in Figure 2(c), this finding suggests that initially randomly oriented 5P-Py molecules align at higher temperatures to bind upright-standing to the ZnO. Both observations, the occurrence of an IGS below $E_F$ as well as the reduction of the work function are consistent with bond formation of the 5P-Py *via* the pyridine end group. The reduced work function can be understood by an alignment of the intrinsic molecular dipole moments along the surface normal and by the formation of interfacial dipoles due to more N–Zn bonds, just as it was the case for pyridine/ZnO(10-10). It should be noted, that both, the comparably small intensity of the 5P-Py IGS as well as the larger work function of 5P-Py compared to pyridine/ZnO suggests an incomplete formation of an upright-standing ML.

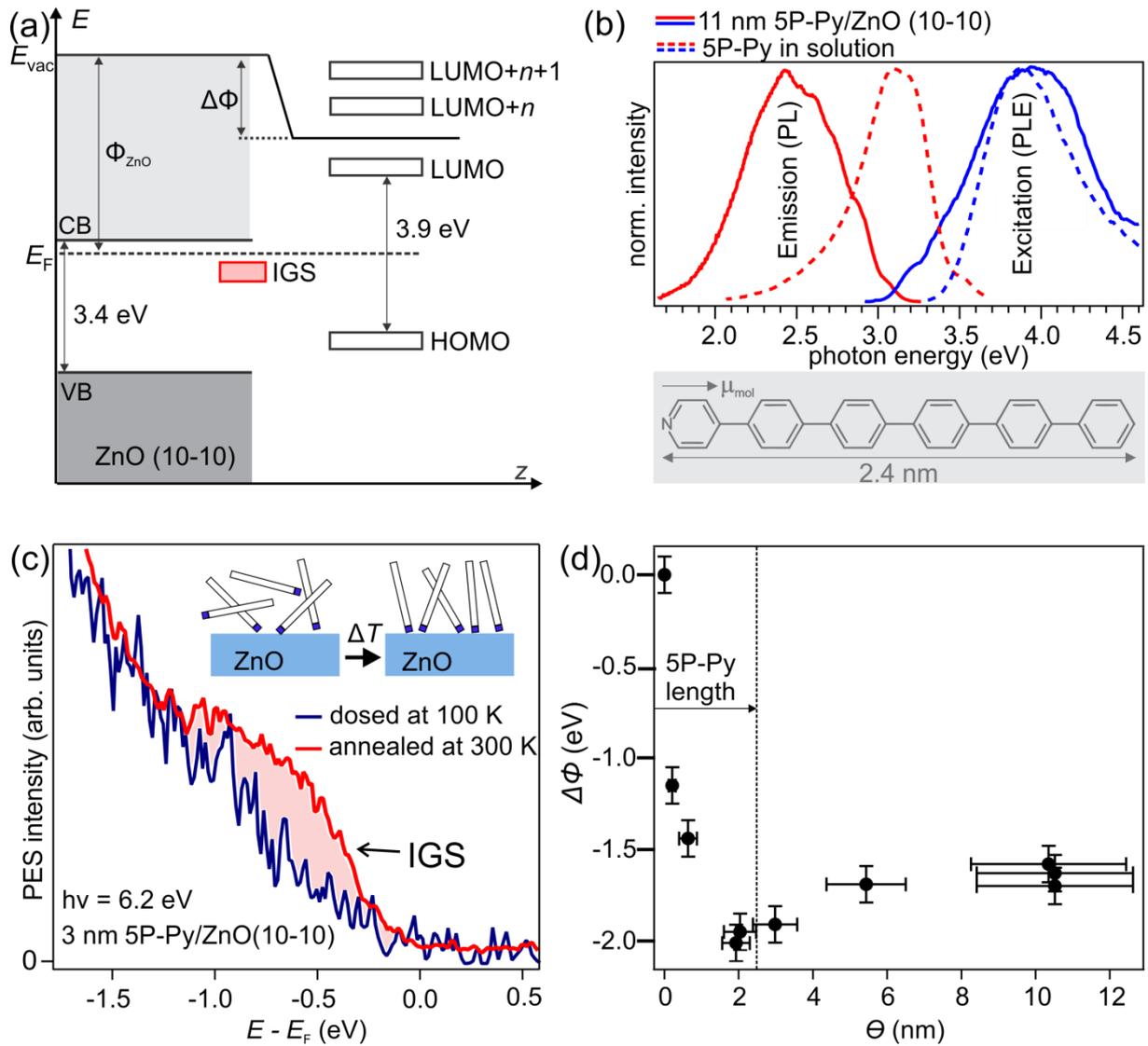

*Figure 2 (a) Schematic of the expected ELA at the 5P-Py/ZnO(10-10) interface. (b) Room temperature PL and PLE of 5P-Py. Inset: 5P-Py molecule. (c) 1PPE of a thin film of 5P-Py/ZnO. An occupied state (IGS) is formed similar to pyridine/ZnO(10-10) upon annealing the as-dosed film to 300 K. Inset: Suggested ordering process of 5P-Py. (d) Work function reduction observed upon deposition of 5P-Py on ZnO at a substrate temperature of T = 300 K. The work function is minimized for nominal coverages of the length of 5P-Py, indicative of the formation of an at least partially upright-standing first ML. The error bars of the measured work function values are estimated values based on the variation of the work function across the sample surface (inhomogeneity of the film) and the uncertainty of the time when the adlayer preparation was started (i.e. the uncertainty of the work function prior to dosing). The error bar of the coverage is estimated from the uncertainty in determining the deposition rate with the QCM.*

We use the change in work function induced by 5P-Py adsorption to determine the completion of the first ML on ZnO(10-10). From now on, all 5P-Py layers are deposited at a substrate temperature of 300 K, in order to provide sufficient thermal energy to the system to equilibrate. Fig. 2(d) shows the work function reduction through 5P-Py adsorption as a function of nominal layer thickness provided by the calibrated QCM. Analogous to the work function reduction observed for pyridine, the work function is first reduced monotonically upon increasing the 5P-Py layer thickness and minimized at a nominal layer thickness of 1.9 nm where ΔΦ = -2.1 eV. For higher 5P-Py coverages, the work function increases again slightly. As indicated in Fig. 2(d), the minimum work function occurs almost exactly for nominal layer thicknesses of the length of the 5P-Py molecule (2.4 nm). This observation strongly suggests that, for low coverages, upright-standing growth of 5P-Py dominates. The smaller work function reduction for 5P-Py compared to pyridine can be rationalized by incomplete upright-ordering of the first ML and possibly smaller charge transfer through N-Zn bond formation from the dye molecule.

The multilayer structure of 5P-Py/ZnO(10-10) is analyzed using XRD. Fig. 3 shows a Θ-2Θ XRD scan as a function of scattering vector $q_z$ for a film with a nominal thickness of 11 nm. Besides the surface reflectivity of the ZnO surface with 5P-Py film at small $q_z$, two strong Bragg reflections can be observed at $q_z$ = 2.24 Å$^{-1}$ and $q_z$ = 1.65 Å$^{-1}$. The first stems from the (100) plane of the ZnO surface. The second peak originates from 5P-Py crystallites with a lattice spacing of $d_{20\bar{3}}$ = 3.82 Å. This value coincides with the lattice spacing of the (20-3) lattice plane in the herringbone structure of 6P[45,46]. In addition to the two intense peaks in the XRD curve, we also observe a reflection at $q_z$ = 1.41 Å$^{-1}$. Under the assumption of similar 6P and 5P-Py unit cells, this reflection can be assigned to the (111) lattice plane of 5P-Py crystallites. The intensity of this (111) reflection is about one order of magnitude smaller than the (20-3) reflection, which indicates that most molecules adopt an orientation, where the (20-3) plane of the 5P-Py crystallites forms the contact plane with the substrate. The (20-3) orientation of crystallites corresponds to flat lying 5P-Py molecules, but also for crystallites with an (111) orientation molecules are nearly flat lying.

We conclude that, in contrast to the (partially) upright standing ML, 5P-Py arranges in a herringbone structure with the long axis of the molecule aligned along the substrate plane for multilayer coverages. It should be noted that the XRD results only determine the orientation of crystalline multilayers and do not exclude the presence of an interfacial layer with different molecular orientation.

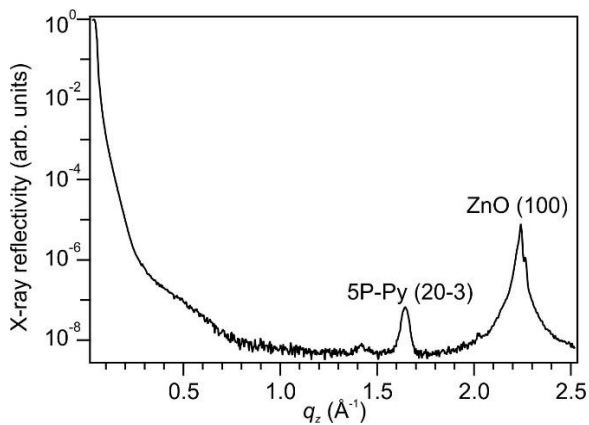

*Figure 3 XRD scan of a 5P-Py film of ~11 nm thickness. The two dominant Bragg reflections can be assigned to the (100) plane of the ZnO substrate surface and the (20-3) plane of 5P-Py in the herringbone structure.*

The growth of 5P-Py multilayer coverages on ZnO(10-10) can be compared to 6P on ZnO(10-10)[17]. There, upon deposition at room temperature and at sub-ML coverage, the coexistence of islands with upright standing 6P and needles of flat lying 6P is observed. At higher coverages, the needles coalesce leading to crystalline films of flat lying 6P. It seems reasonable that due to the preferential binding with the nitrogen lone pair to Zn surface atoms, for 5P-Py the tendency to adsorb in the upright standing geometry is enhanced. It is, however, *a priori* unclear, whether the up-right standing geometry persists for nominal 5P-Py multilayers.

Due to the chemical similarity of 5P-Py and 6P and the permanent dipole moment of 5P-Py aligned in plane with the flat-lying molecule, it seems reasonable to assume a comparable potential step at the vacuum interface for flat-lying 5P-Py and 6P molecules. Yet, we observe a considerably larger work function reduction $\Delta\Phi_{5P-Py}$ = -1.7 eV (cf. Fig. 2(d) for $\Theta$ > 5 nm) compared to $\Delta\Phi_{6P,flat}$ = -0.85 eV for flat-lying 6P films on the same ZnO surface. This strongly suggests the persistence of upright-standing 5P-Py in the first monolayer even for large coverages.

**3.3 Unoccupied states of 5P-Py/ZnO(10-10) and interface sensitivity**

In order to address the unoccupied electronic structure of the o/i interfaces, we performed 2PPE experiments in which bound, normally unoccupied electronic states are transiently populated through optical excitation and then probed in photoemission with a second photon that photoionizes the sample. This process is illustrated by Fig. 4(b) (left), which also shows that a variation of the photon energy should lead to an energy shift of the spectral signature of such intermediate state (i.e. between $E_F$ and $E_{vac}$) by $\Delta h\nu$. In the case of photoemission *via* final states above the vacuum level (cf. Fig. 4(b), right), the spectral feature should remain at the same energy and be independent of the photon energy used in the experiment.

Fig. 4(a) shows an exemplary single-color 2PPE spectrum of 11 nm 5P-Py/ZnO(10-10) measured with a photon energy of hv = 2.52 eV (yellow). The spectrum is plotted versus an intermediate state energy axis $E_{int} - E_F = E_{fin} - h\nu$ (bottom) and a final state energy axis $E_{fin} - E_F = \Phi + E_{kin}$ (top). Besides the secondary electron cut-off, three spectral signatures can be discerned. The spectrum is fitted by three distinct Gaussian peaks (A–C) on top of an exponential secondary electron background (dotted line). The character of the observed spectral signatures A–C can experimentally be identified by measuring their energetic positions as a function of photon energy as discussed above and plotted in Fig. 4(c): For peaks A and B there is no change of energetic position for photon energies between 2.3 and 2.7 eV. This clearly indicates that these signatures originate from final state resonances with energies of 3.35(5) eV (0.58 eV) and 4.05(5) eV (1.28 eV) above $E_F$ (above $E_{vac}$), respectively. In contrast to that, the energetic position of state C changes linearly as a function of energy, with $\Delta E_{fin} = \Delta h\nu$. Thus, signature C originates from a bound, unoccupied electronic state *below* $E_{vac}$. Its maximum relative to $E_F$ lies, according to the linear fit to the photon energy dependence (dashed line in Fig. 4(c)) at 2.10(5) eV. As peak C is the lowest unoccupied state observed by 2PPE, we assign it to the 5P-Py LUMO, while Peak A and B are likely related to LUMO+*n* and LUMO+*n*+1, as illustrated by the energy level diagram in Fig. 2(a). It should be noted that, due to the spectral vicinity of the final state B as well as due to the adjacent resonance condition between IGS and the LUMO, the real energetic position of the LUMO with respect to $E_F$ underlies an uncertainty: In 2PPE, the photoemission intensity depends on the transition matrix elements for the absorption of the two involved photons as well as on the density of initial, intermediate, and final states. The spectral intensity distribution and, thus, energetic position of the peak maximum of C might, therefore, be affected by the near-resonance condition with the IGS and final state B and must not reflect the exact energetic position of the LUMO with respect to $E_F$.

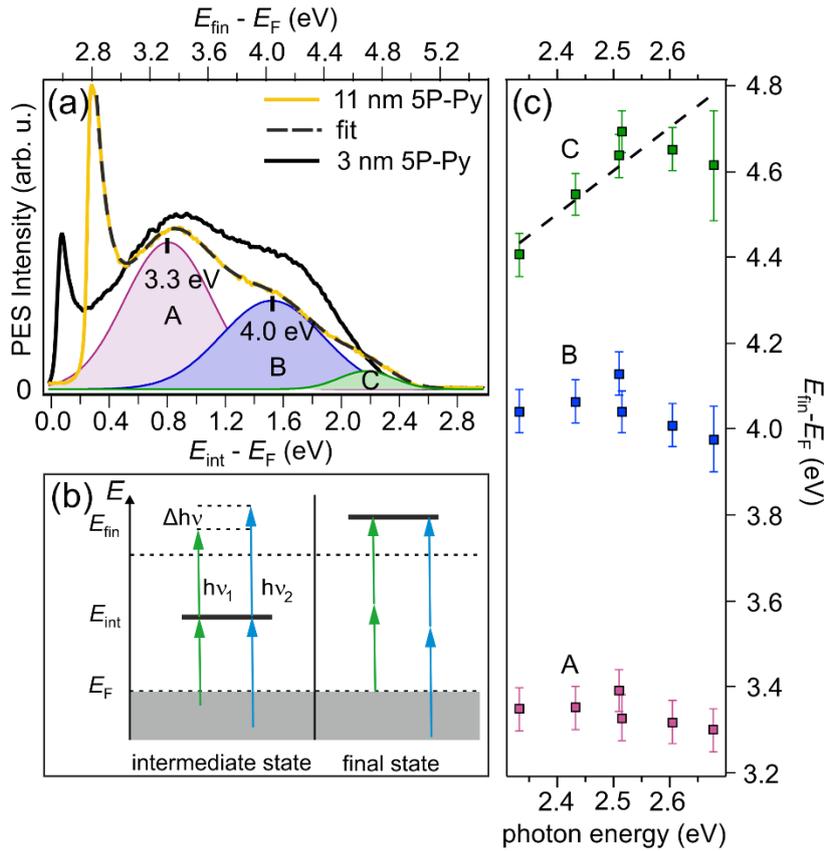

*Figure 4 (a) The single-color (hv = 2.52 eV) 2PPE spectrum of 11 nm 5P-Py/ZnO(10-10) (yellow) exhibits a three peak structure and the low-energy cut-off of the secondary electron background. For 3 nm 5P-Py (black), the third peak is not clearly distinguished. (b) 2PPE photon energy dependence scheme for intermediate (left) and final state (right). (c) Spectral position of peaks A-C as a function of photon energy hv. The dashed line with slope 1 is a guide to the eye.*

While the three-peak structure of the 2PPE spectrum is clearly distinguishable for the nominal thickness of 11 nm 5P-Py/ZnO(10-10), peak C is not discernable in 2PPE spectra of 3 nm 5P-Py (i.e. close to 1 ML), cf. Fig. 4(a) (black). In order to determine whether the ML sample does not expose the 5P-Py LUMO or whether it is energetically shifted and buried underneath final state B, we performed two-color 2PPE, where intermediate states are populated and depopulated by photons of different energy.

Two-color 2PPE spectra for a thin and a thick 5P-Py film on ZnO(10-10) are plotted in Fig. 5(a) versus intermediate state energy (bottom) and final state energy (top). For clarity, we subtracted the uncorrelated photoelectron intensity that is measured when the probe laser pulse reaches the sample picoseconds *before* the pump laser pulse populates normally unoccupied states. Additionally, analogous to Fig. 1(b), the exponential background of secondary electrons was subtracted from the data. Clearly, the 3 nm film (green dashed) exposes an unoccupied state 1.86(5) eV above $E_F$, which we attribute to the 5P-Py LUMO of the first monolayer. Probed by 1.53 eV photons, its spectral signature exhibits a final state energy of 3.4 eV and is, therefore, significantly overlapping with final state A (purple shaded area in Fig. 5(a)).

The diagram in Fig. 5(a) illustrates the population and depopulation process for the single-color (green) and two-color (red) 2PPE experiment. In both cases, the LUMO is populated by photoexcitation with visible light ($h\nu_{VIS}$) from the occupied IGS. Population of the LUMO *via* two photon absorption in the pump pulse can be excluded by excitation density-dependent experiments (not shown) which demonstrate that the photoemission intensity scales linearly with the excitation fluence.[47] As discussed above, in two-color 2PPE, depopulation and, therefore, photoemission occurs in resonance with final state A after the absorption of $h\nu_{IR}$. In the single color experiment on the contrary, photoemission occurs with $h\nu_{VIS}$ = 2.5 eV and in partial resonance with final state B, explaining why C could not be resolved in the black spectrum in Fig. 4(a). We, therefore, conclude that for 3 nm 5P-Py/ZnO(10-10) the observed 2PPE processes of population and depopulation of the LUMO are doubly resonant.

Fig. 5(a) also shows a two-color 2PPE spectrum for a nominal 5P-Py coverage of 11 nm (solid green). It exposes a very similar peak compared to the 3 nm film, which is shifted upwards in energy. Its spectral position at 1.95(5) eV is in agreement with the intermediate state energy of the LUMO of 2.10(5) eV determined by single-color 2PPE. The mismatch between the two experiments can be rationalized when considering the slightly varying resonance conditions with the final states A and B in the two- and single-color experiment, respectively. Also, the peak position may be influenced by the 200 meV higher work function in the case of the 11 nm film that can be determined at the low-energy secondary electron cut-off in Fig. 4(a).

The comparability of the LUMO positions of the monolayer and multilayer samples after interfacial excitation *via* the IGS suggests that, also in the case of the 11 nm sample, a near-surface LUMO of the 5P-Py LUMO is probed. In order to address this question, we compare the 2PPE spectra for both, resonant population of the LUMO from the interfacial IGS and intramolecular, resonant excitation of the 5P-Py molecules at the energy of maximal PLE intensity (3.9 eV, cf. Fig. 2(b)) for 11 nm 5P-Py/ZnO(10-10). Note that in both cases, similar to the resonant photoemission from the LUMO for the thin film, probing with $h\nu_{IR}$ = 1.56 eV and 1.53 eV, respectively, leads to near-resonance conditions with final state A. Remarkably, the LUMO signatures for interfacial (green) and intramolecular photoexcitation (blue) almost coincide, despite the different localization of the photohole at the ZnO surface and in the 5P-Py molecule and despite the fact that interfacial excitation is likely to populate near-surface LUMOs only, while intramolecular excitation can generally occur throughout the organic film. This finding suggests that both experiments probe the LUMO of the same interfacial molecules.

The lifetime of excited states near semiconductor or metal surfaces is determined by the wavefunction overlap with the substrate conduction band. The electron transfer probability in this strong coupling limit therefore depends on the degree of localization of the excited state as well as its distance from the interface[48]. If interfacial and intramolecular excitations lead to the excitation of the same near-surface molecules, they should expose comparable excited state lifetimes. We performed femtosecond time-resolved, two-color 2PPE experiments where we varied the time delay between the pump and the probe laser pulse to monitor the population dynamics of the excited states. Fig. 5(b) shows the transient population of the LUMO for the thin (3 nm) and the thick (11 nm) 5P-Py film after interfacial excitation (green) and compares these to the population dynamics initiated by intramolecular excitation (blue). Clearly, interfacial excitation leads to a fast population decay within the first 100 fs after photoexcitation for the thin (dashed) and thick (solid) film, respectively, while intramolecular excitation causes a significantly longer-lived population of the 5P-Py LUMO that decays on picosecond timescales.

The above observation clearly demonstrates that – despite the spectral coincidence of the LUMO for interfacial and intramolecular excitation – the different excitation schemes populate unoccupied molecular electronic states in different regions of the o/i system and, moreover, that population of the LUMO *via* the IGS provides surface sensitivity even for large coverages of the organic molecule. Based on this, we conclude that, on the one hand, interfacial excitation populates the LUMO near-surface molecules, which is strongly coupled to the ZnO CB, leading to sub-100 fs population decay. Intramolecular excitation throughout multilayer coverages of 5P-Py, on the other hand, shows a significantly longer (average) lifetime, as population decay to ZnO requires exciton diffusion and electron-hole recombination occurs on longer timescales due to momentum conservation. It should be noted that, due to multiple resonance effects in both population and depopulation processes of the LUMO at several photon energies, the exact binding energy of this state within our experimental resolution cannot be determined terminally, however, experimental evidence is sufficient to prove the presence of an unoccupied state at 2.0(3) eV above $E_F$.

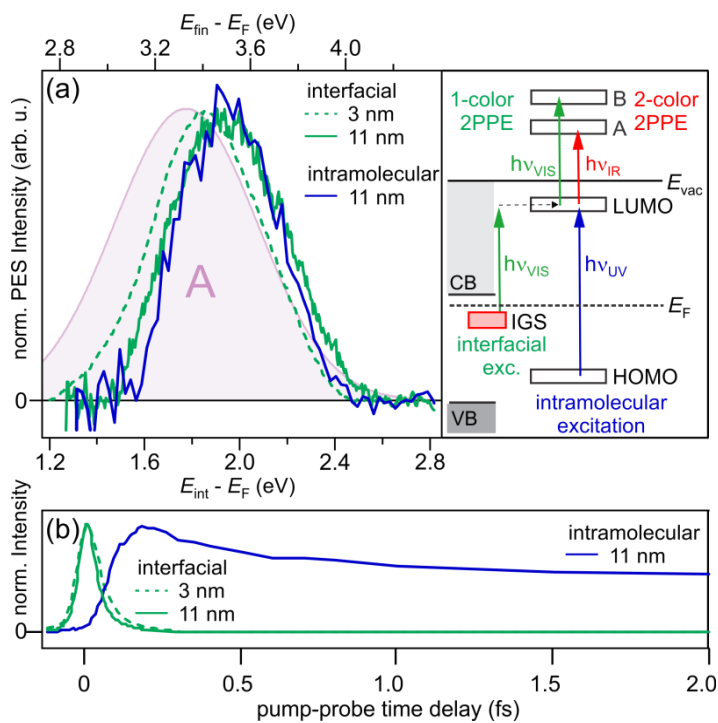

*Figure 5 (a) Background-subtracted, two-color 2PPE spectra from time-resolved pump-probe experiments (integrated signal ± 20 fs around time zero in case of interfacial excitation (green) and from 0 to 200 fs in case of intramolecular excitation). Green dashed: 3 nm 5P-Py/ZnO, $h\nu_{VIS}$ =2.51 eV (pump) and $h\nu_{IR}$ = 1.53 eV (probe). Green solid: 11 nm 5P-Py/ZnO, $h\nu_{VIS}$ =2.52 eV (pump) and $h\nu_{IR}$ = 1.56 eV (probe). Blue solid: 11 nm 5P-Py/ZnO, $h\nu_{UV}$ = 3.90 eV (pump) and $h\nu_{IR}$ = 1.53 eV (probe). Resonance A from the fit in Fig. 4(a) is reproduced for comparison (purple). The diagram illustrates the different population and depopulation pathways. (b) Cross correlation traces of pump and probe laser pulses of the LUMO (integrated 2PPE intensity at ±0.5 eV around peak center) for the same three experiments.*

## 4. Summary and Outlook

We investigated the occupied and unoccupied electronic structure of a model o/i interface: 5P-Py on ZnO(10-10). We found that 5P-Py adsorption leads to a substantial work function reduction of up to -2.1 eV, comparable to the value for adsorption of the pure anchoring group, pyridine, of -2.9 eV reported previously[34]. We showed that both interfaces expose an occupied electronic state below $E_F$, i.e. in the band gap of ZnO, which we interpret as the energetically lowest hybrid exciton at the o/i interface[38] that occurs quasistationarily due to its binding energy *below* the Fermi level of the system. We made use of this IGS to explore the unoccupied electronic band structure of the 5P-Py/ZnO(10-10) interface with one- and two-color 2PPE spectroscopy. This revealed three unoccupied states, two final states above the vacuum level $E_{vac}$ and one bound state 2 eV above the Fermi level $E_F$ that we assign to the 5P-Py LUMO. Furthermore, we show that interfacial excitation of the system using the IGS as an initial state leads to the selective excitation of near-surface molecules even for multilayer coverages of 5P-Py/ZnO(10-10). Their LUMO is strongly coupled to the resonant ZnO CB leading to a rapid population decay on sub-100 fs timescales. Conversely, intramolecular excitation at 3.9 eV results in LUMO population throughout large 5P-Py coverages that is reflected in a significantly enhanced excited state lifetime. In summary, this work does not only provide comprehensive understanding of the occupied and unoccupied electronic structure at a buried o/i hybrid model interface, but also presents a new pathway to disentangle the interfacial properties from bulk responses. The key step in this approach lies in the exploitation of organic anchoring groups and their potential to create interfacial electronic states. Moreover, the interfacial photoexcitation may provide unprecedented insight into the interplay of electronic coupling across o/i interfaces and the electron-hole interaction. Previous studies of charge transfer at such interfaces[5,7,8,9,10,11] used intramolecular photoexcitation and, therefore, created a photohole localized on the molecule that influences the charge transfer probability of the electron due to the attractive Coulomb interaction. Complementarily, the interfacial population of the LUMO produces a photohole at the interface. Control of the photohole position, combined with interface sensitivity will enable the exploration of the effect of the electron-hole interaction on the controversial charge separation dynamics in ZnO-based hybrid solar cells[6,7,10], which will be the topic of a forthcoming publication.

**Acknowledgements**

This work was funded through CRC 951 of the German Science Foundation. We thank Claudia Draxl and Olga Turkina for fruitful discussions. JS is grateful for inspiring exchange with Sabine Reichel. We thank Juan Jesús Velasco-Vélez for his help with AFM measurements for QCM calibration.

# Supporting Information
## to
## Uncovering the (un-)occupied electronic structure of a buried hybrid interface
## (S. Vempati et al.)

**Detailed information on the synthesis of 5P-Py**

**General Synthetic and Analytical Methods and Materials**

Ethyl acetate, tetrahydrofuran (THF), methylene chloride, and ethanol were distilled prior to use. All other starting materials were used as received. Dry solvents were taken from a Pure Solv Micro Solvent Purification System. NMR spectra were recorded on a Bruker DPX 300 Spectrometer (300 MHz for $^1$H and 75 MHz for $^{13}$C) at 25 °C using residual protonated solvent signals as internal standard ($^1$H: $\delta$(CHCl$_3$) = 7.26 ppm; $\delta$(CHDCl$_2$) = 5.32 ppm; $^{13}$C: $\delta$(CDCl$_3$) = 77.16 ppm; $\delta$(CD$_2$Cl$_2$) = 53.84 ppm). IR spectra (neat) were taken on a Bruker VERTEX 70v. For UPLC a Waters UPLC Acquity with a Waters Alliance System (Waters Separations Module 2695, Waters Diode Array Detector 996 and Waters Mass Detector ZQ 2000) was used. For column chromatography silica gel (0.035-0.070 mm, 60 Å pore size) was used. Recycling gel permeation chromatography (GPC) was performed with a JAI LC-9210NEXT using methylene chloride as the eluent. Emission and excitation spectra were taken on a Cary Eclipse Fluorescence Spectrophotometer in CHCl$_3$ at 25 °C, either exciting at 320 nm or observing the emission at 400 nm.

**Synthetic Procedures for Preparation of 5P-Py**

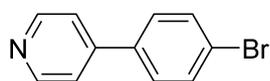
**1-Bromo-4-(4'-pyridyl)benzene.** Under an argon atmosphere 1-bromo-4-iodobenzene (6.6 g, 23.3 mmol), 4-pyridineboronic acid (2.6 g, 21.2 mmol), and aqueous solution of Na$_2$CO$_3$ (2 M, 30 mL) were dissolved in 90 mL of DMF. After several cycles of applying reduced pressure followed by flushing with argon to the mixture, Pd(dppf)Cl$_2$ (0.87 g, 1 mmol) was added and it was stirred at 80 °C overnight. After cooling to room temperature, the mixture was poured on ice and filtered. The precipitate was dissolved in methylene chloride, dried over anhydrous MgSO$_4$, and the solvent evaporated under reduced pressure. The residue was extracted using methylene chloride, dried over anhydrous MgSO$_4$ and evaporated under reduced pressure. Silica gel column chromatography (methylene chloride/methanol = 100/5) of the combined crude fractions afforded 1-bromo-4-(4'-pyridyl)benzene (4.33 g, 18.5 mmol, 87% yield) as a white solid. $^1$H-NMR (300 MHz, CDCl$_3$): $\delta$ [ppm]= 7.51 (m, 4H), 7.63 (m, 2H), 8.69 (m, 2H). $^{13}$C-NMR (75 MHz, CDCl$_3$): $\delta$ [ppm]= 123.9, 128.7, 132.5, 136.9, 147.8, 149.89. ESI-MS [M+H]$^+$ $m/z$ calculated for C$_{11}$H$_9$BrN: 233.991, 235.989; found: 233.975, 235.973.

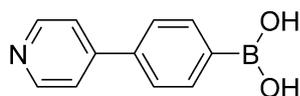
**4-(4'-Pyridyl)phenylboronic acid.** Under an argon atmosphere 1-bromo-4-(4'-pyridyl)benzene (4.1 g, 17.5 mmol) and B(O$i$-Pr)$_3$ (4.9 mL, 21 mmol) were dissolved in 40 mL of dry THF. After cooling to -70 °C, $n$-BuLi (2.2 M in hexanes, 9.6 mL, 21 mmol) was added over 1.5 h and it was stirred for 30 min. The solution was warmed to -20 °C and aqueous HCl was added (1 M, 15 mL). After warming to room temperature, the brown precipitate was filtered and washed with ethanol to yield a brown solid (2.5 g), which was used without further purification.

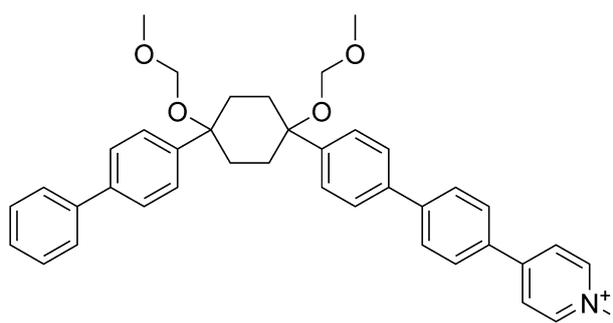 **5P-Py-precursor.** 1-(4-Bromophenyl)-4-(biphenyl-4-yl)-1,4-bis(methoxymethyloxy)cyclohex-ane[1] (750 mg, 1.5 mmol) was dissolved in a mixture of ethanol/THF/DMF (4/4/1, 18 mL) and degassed by several cycles of applying applying reduced pressure followed by flushing with argon. Pd(dppf)Cl$_2$ (40 mg, 0.05 mmol) and Cs$_2$CO$_3$ (1.06 g, 3.26 mmol) were added and the mixture stirred for 15 min. The crude 4-(4'-pyridyl)phenylboronic acid (324 mg, 1.63 mmol) was added and the mixture stirred at 60 °C overnight. After cooling to room temperature, water was added and the mixture was extracted with ethyl acetate. The combined organic layers were washed with saturated aqueous NaCl solution, dried over anhydrous MgSO$_4$ and evaporated under reduced pressure. The crude product was purified by silica gel column chromatography (ethyl acetate) and recycling GPC to yield the product as a white solid (450 mg, 0.77 mmol, 52% yield). $^1$H-NMR (300 MHz, CD$_2$Cl$_2$): δ [ppm]= 2.17 (m, 4H, Cy*H*), 2.37 (m, 4H, Cy*H*), 3.41 (s, 3H, OC*H$_3$*), 3.42 (s, 3H, OC*H$_3$*), 4.47 (s, 2H, OC*H$_2$*), 4.48 (s, 2H, OC*H$_2$*), 7.34 (m, 1H, Ar*H*), 7.43 (m, 2H, Ar*H*), 7.58 (m, 12H, Ar*H*), 7.75 (m, 4H, Ar*H*). $^{13}$C-NMR (75 MHz, CD$_2$Cl$_2$): δ [ppm]= 33.6, 56.3, 78.5, 92.7, 121.9, 127.3, 127.4, 127.5, 127.9, 128.0, 128.1, 128.1, 129.3, 137.4, 139.6, 140.6, 141.0, 141.8, 148.2, 150.6. ESI-MS [M+H]$^+$ *m/z* calculated for C$_{39}$H$_{40}$NO$_4$: 586.295; found: 586.267.

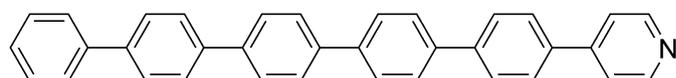 **5P-Py.** The 5P-Py-precursor (518 mg, 0.88 mmol) and NaHSO$_4$ (650 mg, 4.71 mmol) were dissolved in a mixture of DMSO (11 mL) and xylene (36 mL) and stirred at 150 °C for 24 h. After cooling to room temperature, aqueous NaOH solution (1 M, 20 mL) was added and the suspension was stirred for 3 h. Filtration yielded a brown residue, which was thoroughly washed with ethanol and water and then dissolved in 1,2,4-trichlorobenzene to form a clear solution at 215 °C. Air was bubbled through the hot solution for 2 d to complete aromatization. After cooling to room temperature, filtration, and washing with ethyl acetate and methylene chloride, gradient sublimation yielded **5P-Py** as a yellow solid (216 mg, 0.47 mmol, 53% yield). EA calculated for C$_{35}$H$_{25}$N: C: 91.47%, H: 5.48%, N: 3.05%; found: C: 91.08%, H: 5.43%, N: 3.00%. EI-MS [M]$^+$ *m/z* calculated for C$_{35}$H$_{25}$N: 459.198, found: 459.199. IR: ν[cm$^{-1}$]= 3033, 1633, 1589, 1538, 1481, 1449, 1403, 1258, 1227, 1145, 1098, 1072, 1033, 1000, 959, 911, 821, 807, 797, 763, 735.